\documentclass[onecolumn,showpacs,preprintnumbers,showkeys]{revtex4}
\usepackage{amsthm, amscd, amsfonts, amssymb, graphicx}
\usepackage{dcolumn}
\usepackage{bm}
\usepackage[english]{babel}
\usepackage{footnpag}
\usepackage{nccfoots}
\usepackage[symbol*]{footmisc}

\begin{document}

\title{Eliashberg theory with the external pair potential}
\author{Konstantin V. Grigorishin}
\email{konst.phys@gmail.com} \affiliation{Bogolyubov Institute for
Theoretical Physics of the National Academy of Sciences of
Ukraine, 14-b Metrolohichna str. Kiev-03143, Ukraine.}

\begin{abstract}
Based on BCS model with the external pair potential formulated in a work Grigorishin (2017) \cite{grig}, analogous model with electron-phonon coupling and Coulomb coupling is proposed. The generalized Eliashberg equations in the regime of renormalization of the order parameter are obtained. High temperature asymptotics and effect of Coulomb pseudopotential on them are investigated: as in the BCS model the order parameter asymptotically tends to zero as temperature rises, but the accounting of the Coulomb pseudopotential leads to existence of critical temperature. The effective Ginzburg-Landau theory is formulated for such model.

\end{abstract}



\maketitle

\section{Introduction}\label{intr}
\begin{figure}[ht]
\includegraphics[width=8.5cm]{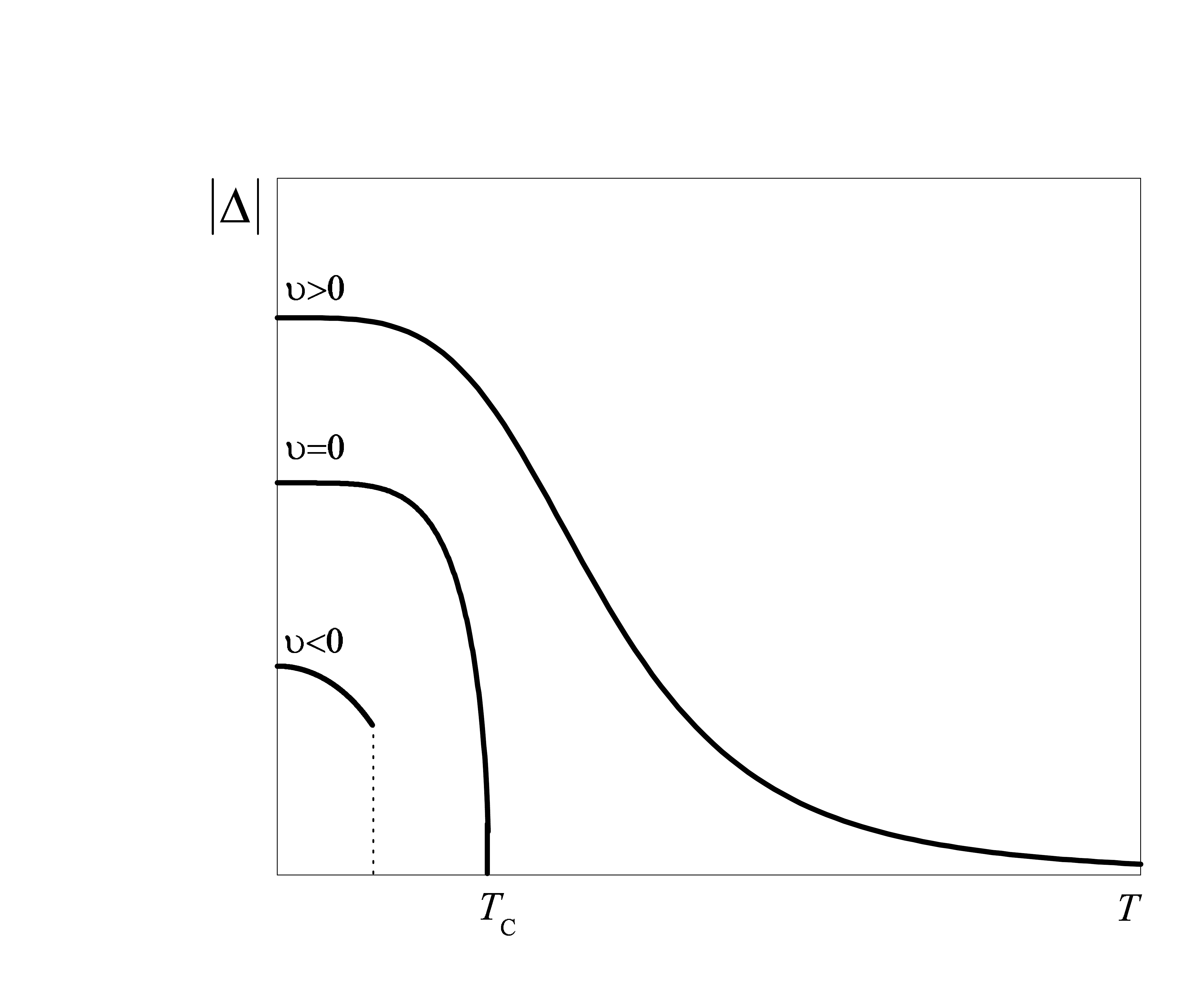}
\caption{Energy gaps $\Delta(T)$ as solutions of Eq.(\ref{1.6}) for three values of the external pair potential $\upsilon$. If $\upsilon<0$ then the pairing of quasiparticles results in increase of the system's energy that suppresses superconductivity and the first order phase transition takes place. If $\upsilon>0$ then the pairing results in decrease of the system's energy and the energy gap $\Delta$ tends to zero asymptotically, that is it does not vanish at any temperature. If $\upsilon=0$ then the second order phase transition superconductor - normal metal takes place.}
\label{Fig1}
\end{figure}

In a work \cite{grig} a hypothetical substance has been considered, where interaction between (within) structural elements of condensed matter (molecules, nanoparticles, clusters, layers, wires etc.) depends on state of Cooper pairs: an additional work $\upsilon$ must be made against this interaction to break a pair. Such a system can be described by BCS Hamiltonian with the external pair potential term. In this model the potential essentially renormalizes the order parameter: if the pairing enlarges energy of the structure then suppression of superconducting (SC) order parameter and the first order phase transition occur, if the pairing lowers energy of the structure then the energy gap is slightly enlarged at zero temperature and at large temperature $T\gg T_{c}$ the gap asymptotically tends to zero as temperature rises:
\begin{equation}\label{1.7}
    |\Delta(T\rightarrow\infty)|=\frac{g\omega\upsilon}{4T},
\end{equation}
where $g$ is a constant of electron-electron coupling via phonons, $\omega$ is the phonon frequency, $\upsilon>0$ is the external pair potential (EPP).
It should be noted that if $g=0$ then for any $\upsilon$ the SC state does not exist ($\Delta=0$ always). This means that only the electron-electron coupling is cause of transition to SC state, but EPP is not. Thus in such system the critical temperature is infinity formally as illustrated in Fig.(\ref{Fig1}). In this model normal $G$ and anomalous $F$ propagators have forms:
\begin{eqnarray}
    G=i\frac{i\varepsilon_{n}+\xi}
    {(i\varepsilon_{n})^{2}-\xi^{2}-|\Delta|^{2}\left(1+\frac{\upsilon}{2|\Delta|}\right)^{2}}\label{1.5a}\\
    F=i\frac{\Delta\left(1+\frac{\upsilon}{2|\Delta|}\right)}
    {(i\varepsilon_{n})^{2}-\xi^{2}-|\Delta|^{2}\left(1+\frac{\upsilon}{2|\Delta|}\right)^{2}},\label{1.5b}
\end{eqnarray}
where $\varepsilon_{n}=\pi T(2n+1)$. The self-consistency condition for the order parameter is
\begin{equation}\label{1.6}
    \Delta=gT\sum_{n=-\infty}^{\infty}\int_{-\omega}^{\omega}d\xi iF(\varepsilon_{n},\xi)
    \Longrightarrow 1=g\int_{-\omega}^{\omega}d\xi
    \frac{1+\frac{\upsilon}{2|\Delta|}}{2\sqrt{\xi^{2}+|\Delta|^{2}\left(1+\frac{\upsilon}{2|\Delta|}\right)^{2}}}
    \tanh\frac{\sqrt{\xi^{2}+|\Delta|^{2}\left(1+\frac{\upsilon}{2|\Delta|}\right)^{2}}}{2T}.
\end{equation}
We can see that the quasiparticle spectrum has a gap $\upsilon/2$ ($\upsilon>0$) even when $\Delta=0$. But this state is not SC because the
ordering $\left\langle aa\right\rangle,\left\langle a^{+}a^{+}\right\rangle$ is absent. Such state can be interpreted as state with a pseudogap due the strong fluctuations of the phase $\phi(\textbf{r},t)$ of the order parameter $\Delta=|\Delta|e^{i\phi}$ so that $\langle e^{i\phi(\textbf{r},t)}\rangle=0$ \cite{grig3}. In a case $\upsilon>0$ at large temperature $T\gg T_{c}$ the gap $\Delta(T)$ tends to zero asymptotically (\ref{1.7}) as temperature rises. Based on bipolaron model of superconductivity \cite{dzum}, in \cite{grig} has been demonstrated that the size of a Cooper pair is much larger than the mean distance between the Cooper pairs (the pairs are strongly overlapped) even for hypothetical room temperatures, that is the Cooper pairs have fermionic nature.

The above results are based on BCS theory, however real
electron-electron interaction is due to the exchange of virtual
phonons. The corresponding potential is described with an expression
\begin{equation}\label{0.1}
    V(\varepsilon_{2}-\varepsilon_{1},\textbf{p}_{2}-\textbf{p}_{1})
    =u^{2}_{ph}\frac{2\omega_{\textbf{p}_{2}-\textbf{p}_{1}}}{(\varepsilon_{2}-\varepsilon_{1})^{2}-\omega_{\textbf{p}_{2}-\textbf{p}_{1}}^{2}},
\end{equation}
where, $u_{ph}$ is an electron-phonon coupling constant,
$\varepsilon_{1},\varepsilon_{2}$ are energetic parameters of
interacting electrons and $\textbf{p}_{2}-\textbf{p}_{1}$ is
transmitted momentum. Since the electrons interact with small
total momentum $\textbf{p}_{2}+\textbf{p}_{1}=0$, then we can
assume that the transmitted momentum is $\sim 2p_{F}$, at the same
time, near the Fermi surface we have
$\varepsilon_{2}\sim\varepsilon_{1}\sim 0$. Therefore the
expression (\ref{0.1}) can be reduced to
\begin{equation}\label{0.2}
    V(\varepsilon_{2}-\varepsilon_{1}=0,\textbf{p}_{2}-\textbf{p}_{1}=2p_{F})
    =-u^{2}_{ph}\frac{2}{\omega_{2p_{F}}}\equiv-\lambda.
\end{equation}
Thus the real interaction is replaced with an effective point attraction, which is nonzero in the layer of width
$2\omega_{2p_{F}}\sim2\omega_{D}$ (Debay frequency) near Fermi surface \cite{sad}. In other words, the BCS approximation
neglects retardation $\varepsilon_{2}-\varepsilon_{1}$ in el.-phon. interaction (the field of lattice deformation is
supposed without inertia). On the contrary, the Eliashberg model \cite{mahan,ginz} is based on the full interaction (\ref{0.1}).
Besides el.-phon. interaction the screened Coulomb interaction $V_{c}$ between electrons takes place
which has width $\sim\epsilon_{F}$. In metals, as a rule, $\lambda<V_{c}$ that corresponds to repulsive electron-electron
interaction, however in such systems the pairing is possible as result of the second order processes which lead to suppression of the direct Coulomb interaction \cite{ginz,kirz}. A stronger condition $\lambda-V_{c}>0$ can occur in nonmetallic superconductors (for example, in alkali-doped fullerides $\texttt{A}_{n}\texttt{C}_{60}$, where competition between the Jahn-Teller coupling and Hund's coupling takes place \cite{han,han1,nom,grig4}).

Our goal is to develop model of superconductivity with EPP using the electron-electron interaction in a form (\ref{0.1}) and
accounting the Coulomb repulsion. In the section \ref{eliashberg} we develop the Eliashberg theory with EPP and investigate high temperature asymptotics in the absence of the Coulomb pseudopotential and in the presence of one. In the section \ref{ginz}, based on effective Ginzburg-Landau theory for the BCS model with EPP developed in \cite{grig}, we formulate the effective GL theory based on the Eliashberg theory.

\section{Generalization of Eliashberg equations}\label{eliashberg}

Let us take into account the fact that the electron-electron
interaction is the result of the exchange of virtual phonons and
of screened Coulomb interaction:
$V_{eff}(\textbf{q},i\omega)=u_{ph}^{2}(\textbf{q})iD(\textbf{q},i\omega)+V_{c}(\textbf{q})$,
where $u_{ph}(\textbf{q})$ is an electron-phonon coupling
parameter,
$D(\textbf{q},i\varepsilon)=\frac{i2\omega(\textbf{q})}{(\varepsilon_{n}-\varepsilon_{m})^{2}+\omega^{2}(\textbf{q})}$
is a phonon propagator. Eliashberg equations, unlike
BCS equations, describe the decrease
of effectiveness of the interaction at phonon energies
$\omega(\textbf{q})\ll T$ (thermal phonons are perceived by
electrons as static impurities \cite{ginz}): as a result $T_{c}
\propto \overline{\omega}\sqrt{g}$ for the el.-phon. model unlike $T_{c}
\propto \overline{\omega} g$ in BCS theory for $g\gg 1$. Moreover, renormalization of electron's mass takes place: $m=m_{0}\left(1+g\right)$ at low temperatures $T\ll \overline{\omega}$, on the contrary at high temperature $T\gg\overline{\omega}$ the renormalization is negligible $m=m_{0}\left(1+\frac{\mathrm{const}}{T^{2}}\right)$ \cite{ginz}.

Like Eqs.(\ref{1.5a},\ref{1.5b}) the normal $G$ and anomalous $F$
propagators take forms:
\begin{eqnarray}
    G=i\frac{i\varepsilon_{n}Z(p)+\widetilde{\xi}}
    {\left(ipZ(p)\right)^{2}-\widetilde{\xi}^{2}-|\widetilde{W}(p)|^{2}}\label{2.1a}\\
    F=i\frac{\widetilde{W}(p)}
    {\left(ipZ(p)\right)^{2}-\widetilde{\xi}^{2}-|\widetilde{W}(p)|^{2}},\label{2.1b}
\end{eqnarray}
where $p\equiv[\textbf{p},ip=i\pi T(2n+1)]$ and
\begin{eqnarray}\label{2.2}
    \widetilde{W}(p)=W(p)+\frac{W(p)}{|W(p)|}\frac{\upsilon}{2}.
   \end{eqnarray}
The self-energies are determined by the self-consistency
conditions:
\begin{eqnarray}
    S(\textbf{p},ip)=-\int\frac{d^{3}q}{(2\pi)^{3}}T\sum_{iq}V_{eff}(\textbf{q},iq)iG(\textbf{p}+\textbf{q},ip+iq)\label{2.3a}\\
    W(\textbf{p},ip)=-\int\frac{d^{3}q}{(2\pi)^{3}}T\sum_{iq}V_{eff}(\textbf{q},iq)iF(\textbf{p}+\textbf{q},ip+iq).\label{2.3b}
\end{eqnarray}
The self energy $S(\textbf{p},ip)$ can be broken up into symmetric
and antisymmetric parts:
$S(p)=S_{e}(\textbf{p},ip)+ipS_{o}(\textbf{p},ip)$, where $S_{e}$
and $S_{o}$ are both even functions of frequency $ip$. Then
renormalization coefficient $Z$ for a single-particle Green
function are $Z(\textbf{p},ip)=1-S_{o}(\textbf{p},ip)$. Accounting
of $S_{e}$ renormalizes the chemical potential only, that does not
influence on the quasiparticles' spectrum. Thus we have
$ip-\xi-S(p)=ipZ(p)-\widetilde{\xi}$, where
$\widetilde{\xi}=\xi+S_{e}(p)$. The functions $Z(p)$ and $W(p)$
are functions of $(\textbf{p},ip)$. For isotropic s-wave
superconductor we can suppose $|\textbf{p}|\approx p_{F}$ and the
main dependence of $G(p)$ and $F(p)$ on $\textbf{p}$ is through
the factor $\widetilde{\xi}\approx\xi$. Then, using method of
\cite{mahan}, Eqs.(\ref{2.3a},\ref{2.3b}) can be written in a form
of Eliashberg equations:
\begin{eqnarray}
    Z(2n+1)=1+\frac{1}{2n+1}\sum_{m\neq n}\widetilde{\Lambda}(2m+1)V(n-m)\label{2.4a}\\
    Z(2n+1)\Delta(2n+1)=\pi T\sum_{m\neq n}\widetilde{\Phi}(2m+1)\left[V(n-m)-\mu_{c}\right],\label{2.4b}
\end{eqnarray}
where $\mu_{c}=V_{c}\nu_{F}$ is a Coulomb pseudopotential ($\nu_{F}$ is electron density on Fermi surface), $V(m)=2\int\omega
d\omega\frac{\alpha^{2}F(\omega)}{\omega^{2}+\varepsilon_{m}^{2}}$
is a phonon interaction. Note that $V(0)=g$, where $g$ is the
dimensionless strength of the electron-phonon interaction. The gap
function is $\Delta(2n+1)=\frac{W(2n+1)}{Z(2n+1)}$,
\begin{eqnarray}
    \widetilde{\Lambda}(2n+1)=\frac{\varepsilon_{n}}{\sqrt{\varepsilon_{n}^{2}+|\widetilde{\Delta}(2n+1)|^{2}}}\label{2.5a}\\
    \widetilde{\Phi}(2n+1)=\frac{\widetilde{\Delta}(2n+1)}{\sqrt{\varepsilon_{n}^{2}+|\widetilde{\Delta}(2n+1)|^{2}}},\label{2.5b}
\end{eqnarray}
where from Eq.(\ref{2.2}) we have
\begin{equation}\label{2.6}
    \widetilde{\Delta}(2n+1)=\Delta(2n+1)+\frac{\upsilon}{2}\frac{\Delta(2n+1)}{|\Delta(2n+1)|}\frac{1}{Z(2n+1)}.
\end{equation}
In Eqs.(\ref{2.4a},\ref{2.4b}) we have excluded the terms with
$m=n$ in the summation, because this term corresponds to elastic
scattering of the quasiparticles on thermal phonons, that does not
make contribution to quasiparticle's mass $m=m_{0}Z$ and to
SC order parameter $\Delta$. Justification of this
fact is given in Appendix \ref{thermal}.

For simplicity we will consider the Einstein model of phonons: all of the phonons
have the same frequency $\omega_{0}$ and
$\alpha^{2}F(\omega)=\omega_{0}g\delta(\omega-\omega_{0})/2$. Then
\begin{eqnarray}\label{2.7}
    &&V(m-n)=\frac{g}{1+\left(\frac{2\pi
    T}{\omega_{0}}\right)^{2}(m-n)^{2}},\label{2.7}\\
    &&Z(2n+1)=1+\frac{2g}{|2n+1|}\sum_{l=1}^{|n|>1}\frac{1}{1+\left(\frac{2\pi
    T}{\omega_{0}}\right)^{2}l^{2}}=Z(-2n-1).\label{2.7a}
\end{eqnarray}
We can see that \textit{efficiency of the el.-el. interaction through phonons exchange decreases with increasing
temperature}. Let us consider particular cases of Eqs.(\ref{2.4a},\ref{2.4b}):

1) $\upsilon=0$, $\mu_{c}=0$. Near $T_{c}$ we have $|\Delta|\ll T$, then Eq.(\ref{2.4b})
takes a form:
\begin{equation}\label{2.8}
Z(2n+1)\Delta(2n+1)=\sum_{m\neq
n}\frac{\Delta(2m+1)}{|2m+1|}V(n-m).
\end{equation}
The asymptotic limit $g\gg 1$ can be found in a simple way.
Assuming that $\frac{2\pi T}{\omega_{0}}$ becomes very large so
that $V(l)$ becomes increasingly small for values $|l|>2$.  The gap equation (\ref{2.8}) can be solved by using only a matrix of dimension two: it is necessary to retain only the gap components $\Delta(1)$ and $\Delta(-1)$ \cite{mahan}, and the renormalization parameter is $Z(1)=Z(-1)=1$. Then:
\begin{eqnarray}\label{2.9}
  \Delta(1)&=&\Delta(-1)V(1) \nonumber\\
  \Delta(-1)&=&\Delta(1)V(-1).
\end{eqnarray}
Setting the determinant equal to zero gives the critical temperature:
\begin{equation}\label{2.10}
T_{c}=\frac{\omega_{0}}{2\pi}\sqrt{g-1}.
\end{equation}

2) $\upsilon>0$, $\mu_{c}=0$. Let temperature is high and the gap is small:
$T\gg \upsilon\gg|\Delta|$, then Eq.(\ref{2.4b}) takes a form:
\begin{equation}\label{2.11a}
Z(2n+1)\Delta(2n+1)=\sum_{m\neq
n}\left[\frac{\Delta(2m+1)}{|2m+1|}V(n-m)+\frac{\upsilon}{2}\frac{\Delta(2m+1)}{|\Delta(2m+1)|Z(2m+1)}\frac{V(n-m)}{|2m+1|}\right].
\end{equation}
Analogously to previous case  we use only a matrix of dimension two:
\begin{eqnarray}\label{2.11b}
  \Delta(1)&=&\Delta(-1)V(1)+\frac{\upsilon}{2}\frac{\Delta(-1)}{|\Delta(-1)|}V(1) \nonumber\\
  \Delta(-1)&=&\Delta(1)V(-1)+\frac{\upsilon}{2}\frac{\Delta(1)}{|\Delta(1)|}V(-1).
\end{eqnarray}
Setting the determinant equal to zero at assumption $|\Delta(1)|=|\Delta(-1)|$ (since the interaction is symmetrical $V(1)=V(-1)$) gives that the energy gap $\Delta$ does not vanish at any temperature:
\begin{equation}\label{2.11}
|\Delta|=\frac{\upsilon}{2}\frac{V(1)}{1-V(1)}\approx\frac{\upsilon
g\omega_{0}^{2}}{8\pi^{2}T^{2}}.
\end{equation}
However, unlike result of BCS theory (\ref{1.7}), the gap tends to
zero faster (as $1/T^{2}$) that is consequence of the
lower effectiveness of the el.-phon. interaction for low phonon
energies $\omega(\textbf{q})\ll T$.

\begin{figure}[ht]
\includegraphics[width=8.5cm]{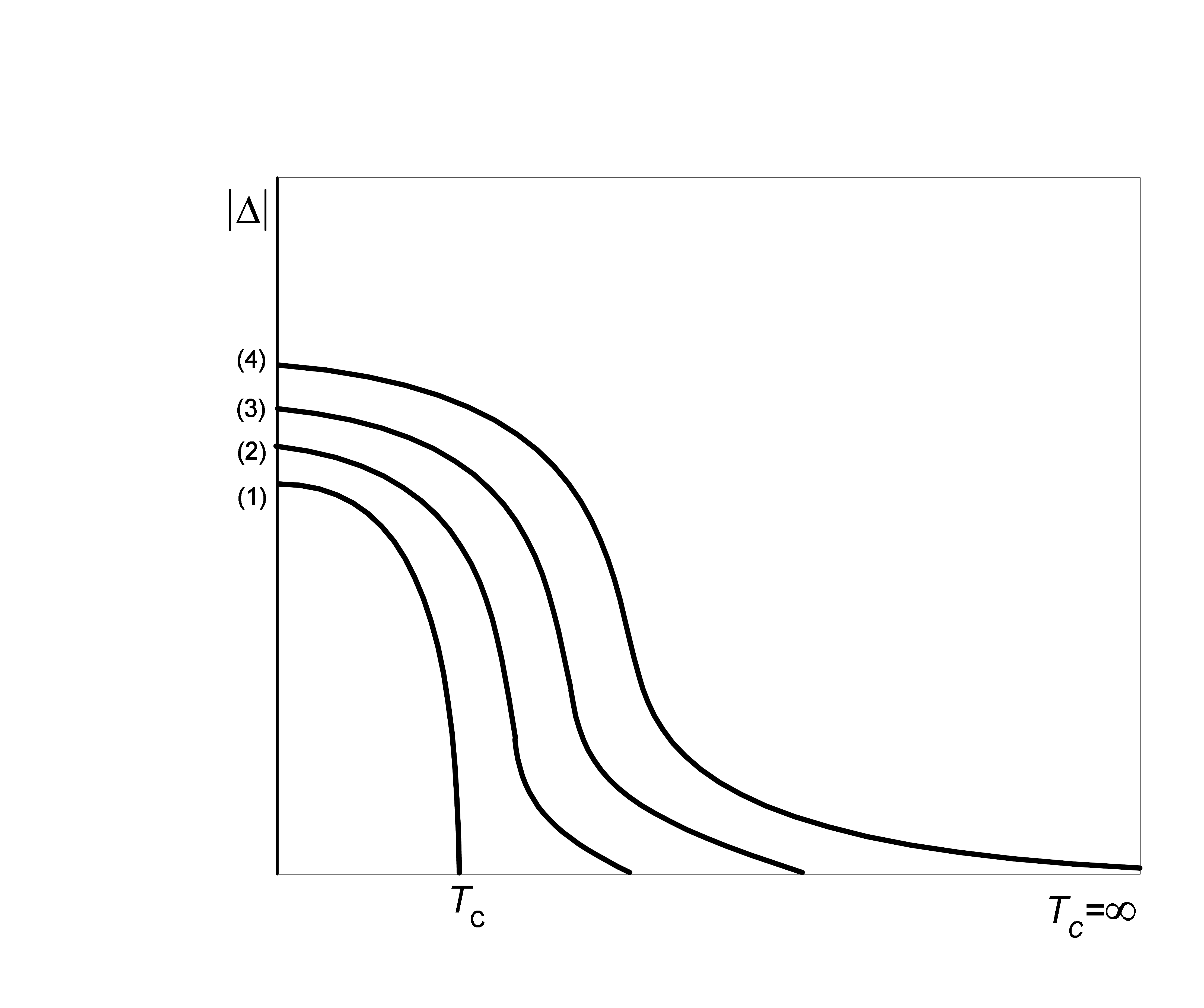}
\caption{Temperature dependencies of the energy gaps: the curve
(1) occures when the external pair potential is absent
$\upsilon=0$ (ordinary BCS or Eliashberg theories); the curves (2)
and (3) are dependencies at $\upsilon>0$ and
$\mu_{c2}>\mu_{c3}\neq 0$; the curve (4) is the dependence when
$\upsilon>0$ but $\mu_{c}=0$, the critical temperature in this
case is equal to infinity because $\Delta\propto 1/T^{2}$.}
\label{Fig2}
\end{figure}

3) $\upsilon=0$, $\mu_{c}\neq 0$. As in previous
consideration we suppose $g\gg 1$. For
$g>\mu_{c}$ the interaction $V(l)-\mu_{c}$ is attractive for small values for $l$
but it becomes repulsive for large values of $l$. For such values
of $m$ that $V(n-m)-\mu_{c}<0$ we suppose $\Delta(2m+1)=0$. Thus, as in previous cases,
it is necessary to retain only the gap components $\Delta(1)$ and
$\Delta(-1)$, then Eq.(\ref{2.4b}) has a form:
\begin{eqnarray}\label{2.12}
  \Delta(1)&=&\Delta(-1)\left[V(1)-\mu_{c}\right] \nonumber\\
  \Delta(-1)&=&\Delta(1)\left[V(-1)-\mu_{c}\right].
\end{eqnarray}
Setting the determinant equal to zero gives critical temperature:
\begin{equation}\label{2.13}
T_{c}=\frac{\omega_{0}}{2\pi}\sqrt{\frac{g}{1+\mu_{c}}-1}.
\end{equation}
Thus it must be $g>1+\mu_{c}$ for such a solution. However in real materials as a rule the relation $g<\mu_{c}$ occurs and the pairing of
electrons can be possible due to reduction of the Coulomb repulsion by Tolmachev's logarithm: $g>\mu_{c}^{*}=\mu_{c}/\left(1+\mu_{c}\ln\frac{\varepsilon_{F}}{\overline{\omega}}\right)$  which is result of the second order processes \cite{ginz,kirz}.

4) $\upsilon>0$, $\mu_{c}\neq 0$, temperature is hight and the gap is small $T\gg \upsilon\gg|\Delta|$. Then Eq.(\ref{2.4b}) takes a
form (we suppose $T\gg\bar{\omega}$ as in previous cases):
\begin{eqnarray}\label{2.14}
  \Delta(1)&=&\Delta(-1)\left[V(1)-\mu_{c}\right]+\frac{\upsilon}{2}\frac{\Delta(-1)}{|\Delta(-1)|}\left[V(1)-\mu_{c}\right] \nonumber\\
  \Delta(-1)&=&\Delta(1)\left[V(-1)-\mu_{c}\right]+\frac{\upsilon}{2}\frac{\Delta(1)}{|\Delta(1)|}\left[V(-1)-\mu_{c}\right].
\end{eqnarray}
From these equations we find the gap like we have done in Eqs.(\ref{2.11b}):
\begin{equation}\label{2.15}
    |\Delta|=\frac{\upsilon}{2}\left[\frac{g}{1+\left(\frac{2\pi
    T}{\omega_{0}}\right)^{2}}-\mu_{c}\right],
\end{equation}
from where we get the critical temperature:
\begin{equation}\label{2.16}
\Delta(T_{c})=0\Longrightarrow
T_{c}=\frac{\omega_{0}}{2\pi}\sqrt{\frac{g}{\mu_{c}}-1}.
\end{equation}
Thus for such a solution it must be $g>\mu_{c}$, that is discussed in Appendix \ref{ratio}. We can see that accounting of the Coulomb pseudopotential leads to existence of critical temperature unlike the result (\ref{2.11}) where the gap tends to zero asymptotically. The critical temperature (\ref{2.16}) is determined by the coupling constants $g,\mu_{c}$ and the frequency $\omega_{0}$ only, like ordinary superconductor (\ref{2.13}), but it does not depend on EPP $\upsilon$. However there is a principal difference of Eq.(\ref{2.16}) from Eq.(\ref{2.13}): if we suppose $\mu_{c}=0$ then $T_{c}(\upsilon=0)\sim\sqrt{g}$ but $T_{c}(\upsilon>0)=\infty$ and the gap (\ref{2.15}) passes into the asymptotic (\ref{2.11}): $|\Delta|\propto 1/T^{2}$.

The expression (\ref{2.15}) can be expanded in the vicinity of
$T_{c}$:
\begin{equation}\label{2.17}
    |\Delta|=\frac{\upsilon g\omega_{0}^{2}}{4\pi^{2}
    T_{c}^{3}}\left(T_{c}-T\right).
\end{equation}
Thus in this model the gap at $T\rightarrow T_{c}$ linearly depends on the temperature difference $\left(T_{c}-T\right)$ unlike ordinary mean field theory without EPP where the temperature dependence of an order parameter is $\left(T_{c}-T\right)^{1/2}$. Temperature dependencies of the
energy gap for different parameters are shown in Fig.\ref{Fig2}.

\section{Effective Ginzburg-Landau theory}\label{ginz}

In a work \cite{grig} the effective Ginzburg-Landau theory for the BCS model with EPP has been formulated. Corresponding free energy
functional has a form:
\begin{equation}\label{3.1}
F_{s}-F_{n}=V\sum_{\textbf{q}}\left[-A|\Delta_{\textbf{q}}|+\frac{B}{2}|\Delta_{\textbf{q}}|^{2}
+C\left(\textbf{q}-2e\textbf{a}_{\textbf{q}}\right)^{2}|\Delta_{\textbf{q}}|
+\frac{1}{2\mu_{0}\hbar^{2}}\left(q^{2}a_{\textbf{q}}^{2}-(\textbf{q}\textbf{a}_{\textbf{q}})^{2}\right)\right],
\end{equation}
where the last term is energy of the magnetic field $\frac{1}{2\mu_{0}}\left(\texttt{curl}\textbf{A}\right)^{2}$, the
coefficients are
\begin{equation}\label{3.2}
A=\nu_{F}g\left(\frac{\hbar\omega_{0}}{2T}\right)^{2}\upsilon,\quad B=\nu_{F}\frac{\omega_{0}}{T},\quad C=\nu_{F}\frac{\omega_{0}}{144T^{3}m^{2}}\upsilon.
\end{equation}
From the free energy functional we can obtain an equilibrium value of the gap, value of the free energy in this point and the
critical momentum of a Cooper pair accordingly:
\begin{eqnarray}
&&\frac{\delta F}{\delta|\Delta|}=0\quad\Longrightarrow\quad
|\Delta|_{min}=\frac{A}{B}\left(1-\frac{C}{A}q^{2}\right)\label{3.3}\\
\nonumber\\
&&(F_{s}-F_{n})_{min}=-\frac{A^{2}}{2B}\left(1-\frac{C}{A}q^{2}\right)^{2}\Longrightarrow
q_{c}^{2}=\frac{A}{C}.\label{3.4}
\end{eqnarray}
If $q=0$ then we obtain Eq.(\ref{1.7}) for the equilibrium value of the gap. Functional (\ref{3.1}) can be written in real space using Fourir transformation, however the functional will have a complicated and inconvenient form due to terms $|\Delta_{\textbf{q}}|$ and $q^{2}|\Delta_{\textbf{q}}|$. Then according to \cite{grig} the functional (\ref{3.1}) can be replaced by an effective GL
functional, which has the same symmetry, the same extremes and the same values in these extremes. The effective GL functional has a
form:
\begin{equation}\label{3.5}
F_{s}-F_{n}=\int\left[-B|\Delta(\textbf{r})|^{2}+\frac{B^{3}}{2A^{2}}|\Delta(\textbf{r})|^{4}
+\frac{BC}{A}|\left(-i\nabla-2e\textbf{A}\right)\Delta(\textbf{r})|^{2}
+\frac{(\texttt{curl}\textbf{A})^{2}}{2\mu_{0}}\right]d^{3}r,
\end{equation}

Unlike BCS theory with EPP, accounting of the Coulomb pseudopotential leads to the gap in a form (\ref{2.15}) that
provides existence of critical temperature (\ref{2.16}). In order to account this facts we should to write the functional
(\ref{3.1}) in a form
\begin{equation}\label{3.6}
F_{s}-F_{n}=V\sum_{\textbf{q}}\left[-A\Xi|\Delta_{\textbf{q}}|+\frac{B}{2}|\Delta_{\textbf{q}}|^{2}
+C\left(\textbf{q}-2e\textbf{a}_{\textbf{q}}\right)^{2}|\Delta_{\textbf{q}}|
+\frac{1}{2\mu_{0}\hbar^{2}}\left(q^{2}a_{\textbf{q}}^{2}-(\textbf{q}\textbf{a}_{\textbf{q}})^{2}\right)\right],
\end{equation}
where the coefficient $\Xi$ is such to obtain the gap
(\ref{2.17}):
\begin{equation}\label{3.7}
    |\Delta(T\rightarrow T_{c})|=\Xi\frac{A}{B}=\Xi\frac{g\omega_{0}\upsilon}{4T_{c}}=\frac{\upsilon g\omega_{0}^{2}}{4\pi^{2}
    T_{c}^{3}}\left(T_{c}-T\right)\Longrightarrow\Xi=\frac{\omega_{0}}{\pi^{2}
    T_{c}^{2}}\left(T_{c}-T\right).
\end{equation}
Then the critical momentum of a pair is
\begin{eqnarray}\label{3.8}
q_{c}=\sqrt{\frac{\Xi A}{C}}\propto\left(T_{c}-T\right)^{1/2}.
\end{eqnarray}
and the gain in free energy (at $q=0$) is
\begin{eqnarray}\label{3.9}
(F_{s}-F_{n})_{min}=-\frac{\Xi^{2}A^{2}}{2B}\propto\upsilon^{2}\left(T_{c}-T\right)^{2}.\label{3.4}
\end{eqnarray}
Thus phase transition to superconducting state in the Eliashberg theory with EPP is the second order phase transition like in ordinary GL theory.  Following the above method we can write the effective GL
functional:
\begin{equation}\label{3.10}
F_{s}-F_{n}=\int\left[\textrm{sgn}(T-T_{c})
B|\Delta(\textbf{r})|^{2}+\frac{B^{3}}{2A^{2}\Xi^{2}}|\Delta(\textbf{r})|^{4}
+\frac{BC}{A|\Xi|}|\left(-i\nabla-2e\textbf{A}\right)\Delta(\textbf{r})|^{2}
+\frac{(\texttt{curl}\textbf{A})^{2}}{2\mu_{0}}\right]d^{3}r,
\end{equation}
Basic characteristics of a superconductor (coherence length $\xi$, magnetic penetration depth $\lambda$, GL parameter $\kappa$,
thermodynamical critical field $H_{cm}$, the first $H_{c1}$ and the second $H_{c2}$ critical fields) for ordinary GL theory at
$T\rightarrow T_{c}$, effective GL theory based on BCS theory with EPP at $T\rightarrow\infty$ and effective GL theory based on
Eliashberg theory with EPP at $T\rightarrow T_{c}$ are presented in the following table:

\begin{center}\label{tab1}
\begin{tabular}{|c|c|c|c|}
  \hline\rule{0cm}{0.5cm}
   & GL theory ($\upsilon=0$) & BCS theory with $\upsilon>0$ & Eliashberg theory with $\upsilon>0$, $\mu_{c}\neq0$ \\
  \hline\rule{0cm}{0.5cm}
  $\xi=\hbar\sqrt{\frac{C}{A\Xi}}\propto$ & $(T_{c}-T)^{-1/2}$ & $\frac{1}{\sqrt{T}}$ & $(T_{c}-T)^{-1/2}$ \\
  \rule{0cm}{0.5cm}
  $\lambda=\frac{\Phi_{0}}{2\sqrt{2}\pi\mu_{0}H_{cm}\xi}\propto$ & $(T_{c}-T)^{-1/2}$ & $\frac{T^{2}}{\upsilon}$ & $\frac{1}{\upsilon}\left(T_{c}-T\right)^{-1/2}$ \\
  \rule{0cm}{0.5cm}
  $\kappa=\frac{\lambda}{\xi}\propto$ & const & $\frac{T^{5/2}}{\upsilon}$ & $\frac{1}{\upsilon}$ \\
  \rule{0cm}{0.5cm}
  $H_{cm}=\frac{\Xi A}{\sqrt{\mu_{0}B}}\propto$ & $(T_{c}-T)$ & $\frac{\upsilon}{T^{3/2}}$ & $\upsilon\left(T_{c}-T\right)$ \\
  \rule{0cm}{0.5cm}
  $H_{c1}=\frac{\Phi_{0}}{4\pi\mu_{0}\lambda^{2}}\ln\kappa\propto$ & $(T_{c}-T)$ & $\frac{\upsilon^{2}}{T^{4}}$ & $\upsilon^{2}\left(T_{c}-T\right)$ \\
  \rule{0cm}{0.5cm}
  $H_{c2}=\frac{\Phi_{0}}{2\pi\mu_{0}\xi^{2}}=\sqrt{2}\kappa H_{cm}\propto$ & $(T_{c}-T)$ & $T$ & $\left(T_{c}-T\right)$\\
  \hline
\end{tabular}
\end{center}
We can see from the table that in the effective GL theory based on the Eliasberg approach the temperature dependencies of the basic
characteristics of a superconductor is similar to the ordinary GL theory unlike the approach based on BCS theory. In particular,
this model restores ordinary temperature dependence of the coherence length $(T_{c}-T)^{-1/2}$ after the BCS model with EPP
where it decreases as $1/\sqrt{T}$ at large $T$ that corresponds extremely small value of order of interatomic distances. However
the effective free energy functional (\ref{3.10}) has an extraordinary form due temperature dependence of the gap as
$\Delta\propto(T_{c}-T)$ unlike the ordinary GL theory where $\Delta\propto(T_{c}-T)^{1/2}$. We cannot write the effective
functional as $-(T_{c}-T)^{2}a\Delta^{2}+b\Delta^{4}$ (that gives the desired temperature dependence of the gap too) because we will
obtain the free energy as $F_{min}\propto(T_{c}-T)^{4}$, that means the phase transition to superconducting state will not be
second-order phase transition.

\section{Summary}\label{result}

Based on BCS model with EPP formulated in \cite{grig} in this work we have proposed analogous model with electron-phonon coupling and Coulomb coupling. We have obtained the generalized Eliashberg equations (\ref{2.4a},\ref{2.4b},\ref{2.5a},\ref{2.5b},\ref{2.6}) for the case of the external pair potential. Only electron-electron coupling is the cause of the SC ordering, but not EPP, however the potential essentially renormalizes the order parameter. Solving these equations for the case $\upsilon>0$ (that is the pairing lowers the energy of the molecular structure, that supports superconductivity) we have obtained the following asymptotic solutions.

If electron-phonon interaction is present only (the Coulomb pseudopotential is absent $\mu_{c}=0$) then the energy gap $\Delta$ does not vanish at any temperature, however the gap tends to zero faster (as $1/T^{2}$ - Eq.(\ref{2.11})) than the result of BCS theory (as $1/T$ - Eq.(\ref{1.7})) that is consequence of decreasing of efficiency of the el.-el. interaction through phonons exchange with increasing temperature. On the other hand the accounting of the Coulomb pseudopotential $\mu_{c}\neq 0$ leads to existence of critical temperature (\ref{2.16}), which is much higher than one in pure material. The gap at $T\rightarrow T_{c}$ linearly depends on temperature difference $T_{c}-T$ - Eq.(\ref{2.17}), unlike the ordinary mean field theory without EPP where the temperature dependence of the order parameter is $\left(T_{c}-T\right)^{1/2}$. It should be noted that equilibrium uncorrelated pairs are present at $T>T_{c}$ even, unlike ordinary superconductors (which are described by BCS and Eliashberg theories).

Based on a free energy functional for BCS model with EPP obtained in \cite{grig} we have written free energy functional (\ref{3.6}) for our model which takes into account above-mentioned critical temperature and the linear dependency of the order parameter on the temperature difference $T_{c}-T$. Following \cite{grig} we have obtained the effective Ginzburg-Landau functional, which has the same symmetry, the same extremes and the same values in these extremes as in the initial functional. The temperature dependencies near $T_{c}$ of the basic characteristics of a
superconductor (coherence length, magnetic penetration depth, GL parameter, thermodynamical critical field, the first and the second critical fields) recovers to the temperature dependencies as in the ordinary GL theory after the BCS model with EPP.

\appendix
\section{Scattering on thermal phonons}\label{thermal}

Let us consider Eq.(\ref{2.4a}) with the symmetrical term $m=n$ and when $\Delta=0$, $\upsilon=0$:
\begin{eqnarray}
    Z(2n+1)=1+\frac{1}{2n+1}\sum_{m=-\infty}^{+\infty}\frac{\varepsilon_{m}}{|\varepsilon_{m}|}V(n-m)\label{A1}
\end{eqnarray}
Then we have
\begin{eqnarray}
    Z(1)=Z(-1)=1+V(0),\quad Z(3)=Z(-3)=1+\frac{1}{3}\left[V(0)+2V(1)\right],\ldots,\label{A2}
\end{eqnarray}
where $V(0)=2\int\omega d\omega\frac{\alpha^{2}F(\omega)}{\omega^{2}}\equiv g$, $V(l)=2\int\omega d\omega\frac{\alpha^{2}F(\omega)}{\omega^{2}+(2\pi Tl)^{2}}$. We can see that at $T\rightarrow\infty$ (this means
$T\gg\overline{\omega}$) we have $V(l\neq 0)\rightarrow0$. Thus $Z(T\rightarrow\infty)=1+g$. However it must be $Z(T\rightarrow
0)=1+g$ and $Z(T\rightarrow\infty)=1$ \cite{ginz}: electron's mass $m_{0}$ is renormalized due el.-ph. interaction as $m=m_{0}Z$ (an
electron is being followed by cloud of virtual phonons), but at high temperatures ($T\gg\overline{\omega}$) the renormalization is
negligible, that underlies the experimental method of finding of the constants $g$.

\begin{figure}[ht]
\includegraphics[width=8.0cm]{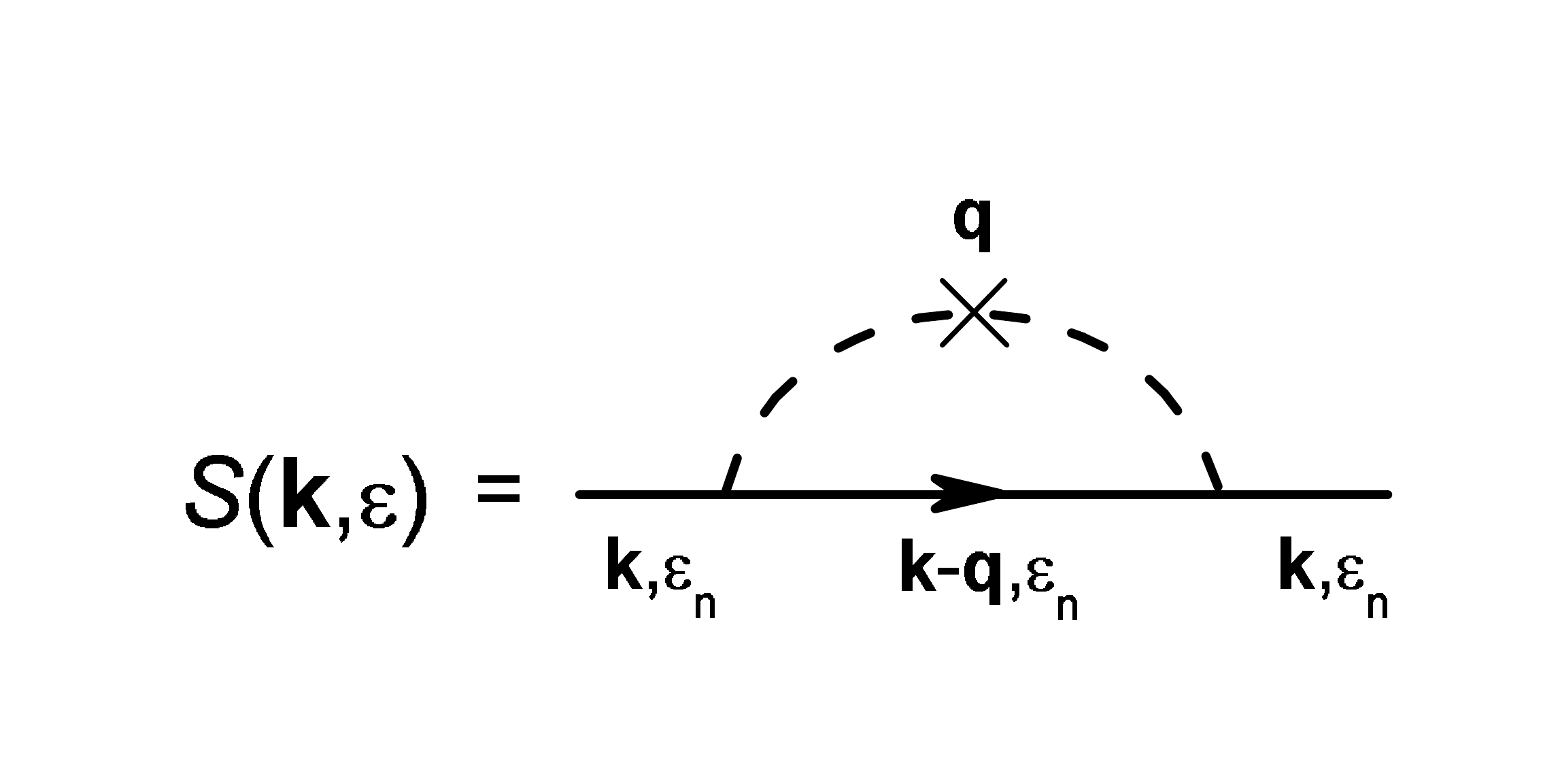}
\caption{Second order phase diagram for elastic scattering of an electron on averaged field of
impurities \cite{sad}.} \label{Fig1A}
\end{figure}

Let us consider the term with $m=n$ in Eqs.(\ref{2.3a},\ref{A1}). This term corresponds to elastic interaction because the energetic
parameters of electron and phonon do not change $\varepsilon_{n}=\varepsilon_{m}$ but the momentum changes as
$\textbf{p}\rightarrow\textbf{p}-\textbf{q}$. Let us consider elastic scattering of an electron on impurities of concentration
$\rho$ using diagrammatics for disordered systems \cite{sad} - Fig.(\ref{Fig1A}). The self-energy has a form:
\begin{equation}\label{A3}
S(\textbf{k},\varepsilon_{n})=-\rho
U^{2}\nu_{F}\int_{-\infty}^{+\infty}\frac{i\widetilde{\varepsilon}_{n}+\xi}
{\widetilde{\varepsilon}_{n}^{2}+\xi^{2}}d\xi=-i\frac{\varepsilon_{n}}{|\varepsilon_{n}|}\pi\rho
U^{2}\nu_{F}\equiv-i\gamma \textrm{sgn}\varepsilon_{n}.
\end{equation}
Thus the elastic impurities do not influence upon effective mass of quasi-particles but they stipulate the damping of quasi-particles $\gamma\textrm{sgn}\varepsilon_{n}$ (the mean free time and the free length are determined as $\tau=\frac{1}{2\gamma},\quad l=v_{F}\tau$). The self-energy (\ref{2.3a}) with the symmetrical term $m=n$ only, using (\ref{A1}), takes a form: \begin{equation}\label{A4}
S=i\varepsilon_{n}S_{o}=i\varepsilon_{n}(1-Z)= -i\varepsilon_{n}\frac{1}{2n+1}\frac{\varepsilon_{n}}{|\varepsilon_{n}|}V(0)=-i\frac{\varepsilon_{n}}{|\varepsilon_{n}|}\pi Tg.
\end{equation}
Comparing Eq.(\ref{A4}) and Eq.(\ref{A3}) we can see that elastic scattering of the quasiparticles on thermal phonons is equivalent to the elastic scattering on impurities, and it does not influence upon effective mass of quasi-particles but stipulates the damping of quasi-particles. Hence the term with $m=n$ must be omitted in the equation for $Z$.

Let us consider Eq.(\ref{2.4b}) when $T\gg \upsilon\gg|\Delta|$ (the first term can be omitted):
\begin{eqnarray}
    Z(2n+1)\Delta(2n+1)=\sum_{m=-\infty}^{+\infty}\frac{\upsilon}{2}\frac{\Delta(2m+1)}{|\Delta(2m+1)|}\frac{V(n-m)-\mu_{c}}{Z(2m+1)|2m+1|}.\label{A5}
\end{eqnarray}
For $m=n$ we have nonzero gap at $T\rightarrow\infty$:
\begin{eqnarray}
    |\Delta(2n+1)|=\frac{\upsilon}{2}\frac{g-\mu_{c}}{|2n+1|Z^{2}(2n+1)}.\label{A6}
\end{eqnarray}
This gap, like the renormalization factor $Z$ at $m=n$, is result of scattering on thermal phonons. Above we have seen that this scattering stipulates the damping of quasi-particles but it does not make contribution to the effective mass. It can be assumed, that this scattering cannot lead to coherent assemble of Cooper pairs, so that for the gap (\ref{A6}) we have $\left\langle\Delta(\textbf{r},t)\right\rangle=|\Delta(\textbf{r})|\left\langle e^{i\phi(\textbf{r},t)}\right\rangle=0$ although it can be
$|\Delta(\textbf{r})|\neq0$, i.e. the superconducting ordering is destroyed by phase fluctuations as described in a work \cite{grig3}. Thus the term with $m=n$ must be omitted in the equation for $\Delta$.

\section{The ratio between $g$ and $\mu_{c}$}\label{ratio}

The electron-electron interaction in a metal has a form \cite{ginz}:
\begin{equation}\label{B0}
  V_{eff}(q,\omega)=\frac{4\pi e^{2}}{q^{2}\varepsilon(q,\omega)}+\frac{\lambda^{2}(q)}{\varepsilon^{2}(q,\omega)}\frac{\Omega^{2}}{\omega^{2}-\omega^{2}(q)},
\end{equation}
where $\varepsilon(q,\omega)$ is a dielectric permittivity, $\omega(q)$ is frequency of remormalized phonons $\omega^{2}(q)\approx\Omega^{2}/\varepsilon(q,0)$, $\Omega^{2}=\frac{4\pi n(Ze)^{2}}{M}$ is frequency of bare phonons in the jelly model (plasmons in gas of ions of charge $Ze$, mass $M$ and with concentration $n$). The first term is a screened Coulomb interaction, the second term is an interaction via phonons. The renormalized el.-phon. coupling constant $g$ is connected with the bare constant $\lambda^{2}(q)$ by a following expression:
\begin{equation}\label{B1}
  g=\nu_{F}\int_{0}^{2k_{F}}\frac{qdq}{2k_{F}}\frac{\lambda^{2}(q)}{\varepsilon^{2}(q,0)}\frac{\Omega^{2}}{\omega^{2}(q)}
  =\nu_{F}\int_{0}^{2k_{F}}\frac{qdq}{2k_{F}}\frac{\lambda^{2}(q)}{\varepsilon(q,0)},
\end{equation}
The Coulomb coupling constant $\mu_{c}$ is
\begin{equation}\label{B2}
  \mu_{c}=\nu_{F}\int_{0}^{2k_{F}}\frac{qdq}{2k_{F}}\frac{4\pi e^{2}}{q^{2}\varepsilon(q,0)}.
\end{equation}
Comparing these equations we can see that to be $g>\mu_{c}$ it is necessary the bare el.-phon. interaction exceeds the bare Coulomb interaction $V_{c}(q)=\frac{4\pi e^{2}}{q^{2}}$:
\begin{equation}\label{B3}
  \frac{\lambda^{2}(q)}{V_{c}(q)}>1.
\end{equation}
In the jelly model $\lim_{q\rightarrow 0}\frac{\lambda^{2}(q)}{V_{c}(q)}=1$ \cite{ginz,schr}. This means that $g>\mu_{c}$ can occur without Tolmachev's reduction $\mu_{c}^{\ast}=\mu_{c}/\left(1+\mu_{c}\ln\left(\frac{\varepsilon_{F}}{\omega_{0}}\right)\right)\ll\mu_{c}$.
Thus SC phase can be in the materials with narrow conduction band $\varepsilon_{F}\sim\omega_{0}$, for example, in
alkali-doped fullerides \cite{han,han1,nom,grig4}.

\acknowledgments

This research was supported by theme grant of department of physics and astronomy of NAS of Ukraine: "Noise-inducing dynamics and correlations in nonequilibrium systems", N 0120U101347.

\end{document}